\documentclass[twocolumn,aps,prl,groupedaddress,amssymb]{revtex4}
\usepackage{natbib}
\usepackage{graphicx}
\usepackage{units}

\begin{document}
\hyphenation{Ryd-berg}


\title{Rydberg excitation of Bose-Einstein condensates}


\author{Rolf Heidemann}
\email[Electronic address: ]{r.heidemann@physik.uni-stuttgart.de}
\author{Ulrich Raitzsch}
\author{Vera Bendkowsky}
\author{Bj\"{o}rn Butscher}
\author{Robert L\"{o}w}
\affiliation{5. Physikalisches Institut, Universit\"{a}t
Stuttgart, Pfaffenwaldring 57, 70569 Stuttgart, Germany}

\author{Tilman Pfau}
\email[Electronic address:]{t.pfau@physik.uni-stuttgart.de}
\affiliation{5. Physikalisches Institut, Universit\"{a}t
Stuttgart, Pfaffenwaldring 57, 70569 Stuttgart, Germany}

\date{29 October 2007}

\begin{abstract}
  Rydberg atoms provide a wide range of possibilities to tailor
  interactions in a quantum gas. Here we report on Rydberg excitation
  of Bose-Einstein condensed $^{87}$Rb atoms. The Rydberg fraction was
  investigated for various excitation times and temperatures above and
  below the condensation temperature. The excitation is locally
  blocked by the van der Waals interaction between Rydberg atoms to a
  density-dependent limit. Therefore the abrupt change of the thermal atomic
  density distribution to the characteristic bimodal distribution upon
  condensation could be observed in the Rydberg fraction. The
  observed features are reproduced by a simulation based on local
  collective Rydberg excitations.

\end{abstract}

\maketitle



The ability to tailor interactions among atoms in a quantum gas is
one of the main strengths of ultracold atoms as model systems for
other research fields such as condensed matter physics. So far mostly
effective contact interactions have been used which can be changed
over a wide range in strength and even in sign. Two years ago, the
long-range magnetic dipole-dipole interaction was observed between
atoms in a $^{52}$Cr-BEC \cite{Stuhler:2005}. It was proposed to
induce dynamic dipole-dipole interactions \cite{Low:2005} or a
$1/r$ interaction potential \cite{O'Dell:2000} between ultracold
atoms by light. To induce a static electric dipole-dipole
interaction, it was suggested to admix
 a Rydberg state with a static electric dipole
moment to the atomic ground state by photo-excitation
\cite{Santos:2000}.

Rydberg atoms and ground state atoms are predicted to form unusual
weakly bound molecular states \cite{Greene:2000,Greene:2006}. As the
average interparticle distances in BECs are of the same order of
magnitude as the molecular binding distances, a BEC seems to be the
ideal starting point for photoassociation of these molecules.

A Rydberg atom also constitutes an impurity in the BEC related to
Ref.\,\cite{Chikkatur:2000}, which could be manipulated by electric
fields. The interaction could lead to an agglomeration of ground state
atoms around the impurity as was calculated for ionic impurities
\cite{Massignan:2005}.

As the nature and strength of the interaction among Rydberg atoms can
be tailored by electric fields and microwave fields e.g. from van der
Waals type to dipole dipole type \cite{Vogt:2007} or an isotropic
$1/r^3$ potential \cite{Buchler:2007} they provide a new tool for many
body quantum physics.

In this Letter we describe the excitation of atoms to a Rydberg state
while undergoing a phase transition from a thermal gas to a
Bose-Einstein condensate. We present a model based on collective
excited states to simulate the observed Rydberg fraction across the phase
transition which agrees qualitatively with the observed data.

Using the setup described in \cite{Loew:2007}, we magnetically trap
$^{87}$Rb-atoms in the 5S$_{1/2}, F=2, m_F=2$ state and produce
samples from thermal clouds to Bose-Einstein condensates by means of
forced RF-evaporation. We vary the temperature from \unit[5]{$\mu$K}
down to \unit[200]{nK} crossing the condensation temperature at around
\unit[700]{nK}. After this preparation, the atoms are subject to a
two-photon Rydberg excitation via the 5P$_{3/2}$ state to the
43S$_{1/2}$ state. The duration of the square pulses of excitation
light was varied in this experiment between \unit[170]{ns} and
\unit[2]{$\mu$s}. To avoid significant absorption and heating due to
spontaneous photon scattering, the light is blue detuned by $\Delta$=
\unit[483]{MHz} from the 5$P_{3/2}, F=3$ level. Thus only one photon
per 100 atoms is scattered for the longest excitation time. For the
following experiments the Rabi frequency $\Omega_1$ on the 5S-5P
transition is \unit[11]{MHz}. A Rabi frequency $\Omega_2$ for the
5P-43S transition of \unit[9.7]{MHz} \cite{CommentRabifrequency:2007}
results in a two-photon Rabi frequency $\Omega_1\Omega_2/(2\Delta)$ of
\unit[110]{kHz}. The thermal motion of the atoms is negligible on the
$\mu$s-time scale of the experiments, but interactions among the
Rydberg atoms can lead to collisions and ionisation. By choosing a
Rydberg state with repulsive interactions and applying an electric
extraction field, the unwanted effects of ions are avoided
\cite{Heidemann:2007}.

\begin{figure}
\includegraphics[width=60mm]{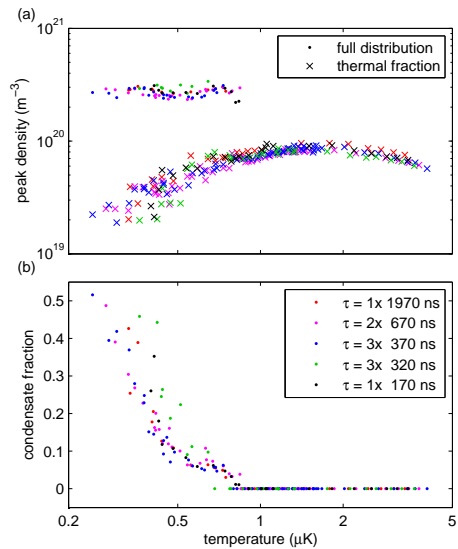}
\caption{\label{Fig_density_ratio} (a) Dependence of the peak density
  of ground state atoms on the temperature. The crosses ($\times$)
  indicate the peak density of the thermal component, the dots
  ($\cdot$) the peak density of the bimodal distribution (where a BEC
  component exists). In (b) the condensate fraction is plotted versus
  the temperature. The colors indicate the length of the Rydberg
  excitation pulse, which was applied in trap before taking the
  absorption picture after a time of flight of 20 ms. The number of
  identical excitations within one sample is denoted by the multiplier
  in the legend. Note that no systematic effect of different
  excitation times on the distributions is visible. The same samples
  delivered the data presented in Fig.\,~\ref{Fig_Rydberg_fraction}(a).}
\end{figure}

The Rydberg excitation is followed immediately by a field-ionisation
of the Rydberg atoms and their detection on a microchannel plate. Up
to three such sequences of excitation and detection are applied to one
sample followed by absorption imaging of the remaining ground state
atoms. From these calibrated detection methods, the total atom number
$N_g$ and the fraction $f=N_R/N_g$ of atoms in the Rydberg state are
derived. This normalisation of the Rydberg atom number $N_R$ is useful
since the total atom number drops significantly from $10^7$ to $10^5$
while reducing the temperature. The temperature of each sample is
derived from the size on the absorption image after an expansion
during \unit[20]{ms} time of flight. The thermal clouds are fitted
using a Bose-distribution neglecting interaction among the ground
state atoms. The BEC component is fitted with a Thomas-Fermi
distribution.  From the parameters of the fit and the known trapping
potential, the in-trap density distributions are calculated and used
for simulating the Rydberg excitation. The peak density and the
condensate fraction are plotted versus temperature in
Fig.\,\ref{Fig_density_ratio} where no effect of the different
excitation times on the atomic distribution is visible. The
bimodality of the density distribution will prove to be crucial for
the understanding of the results of the Rydberg excitation. When the
temperature is decreased below the critical temperature, the central
density increases abruptly by a factor of about 4. When the
temperature is reduced further, more atoms are condensed into the
small constricted region of the BEC reducing the density of the
thermal component. In the condensate the average atomic distance to
the next neighbour is \unit[150]{nm} ($\approx 0.55
n_g^{-\nicefrac{1}{3}}$ \cite{Clark:1954}) which is calculated from
the peak atomic density $n_g$.

\begin{figure}
\includegraphics[width=86mm]{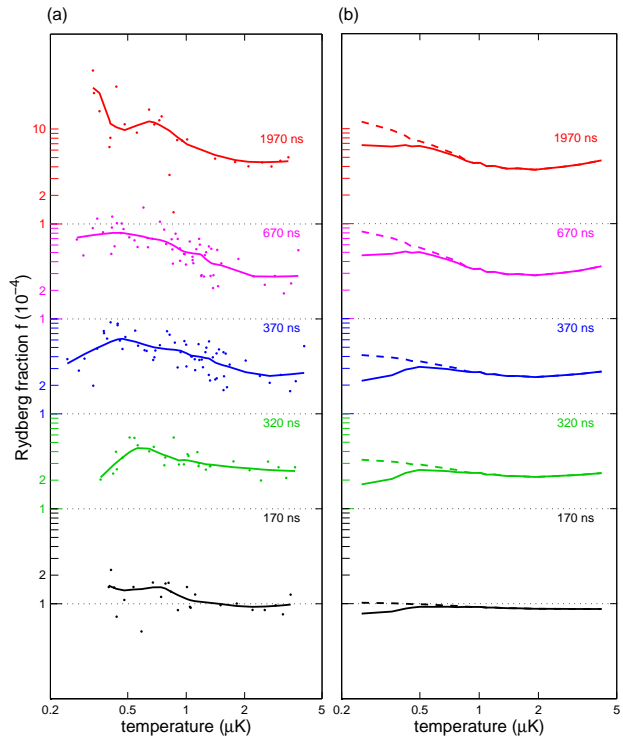}
\caption{\label{Fig_Rydberg_fraction} Rydberg fraction as a function
  of temperature for different excitation times. The data sets for
  each excitation time are color-coded and for clarity shifted such
  that the dotted horizontal lines each indicate $10^{-4}$.  The
  measurements in (a) are smoothed for each excitation time giving the
  solid lines as a guide to the eye \cite{commentlowess:2007}. The
  solid lines in (b) show the calculated Rydberg fractions based on
  the superatom model. The dashed lines show the result of the
  simulation taking only the thermal part of the atomic cloud. This
  confirms the Bose-Einstein condensation as the origin of the kink in
  the Rydberg fraction.}
\end{figure}

In a recent work, we investigated the density dependence of the
Rydberg excitation dynamics in magnetically trapped clouds above the
condensation temperature \cite{Heidemann:2007}. The Rydberg atom
number as a function of excitation time initially increases linearly
with a slope $R$ and saturates to a value $N_{\text{sat}}$ after a
time $\tau_{s}=N_{\text{sat}}/R$. From this previous experimental
observation we know that the saturation is reached faster with
increasing density and that the saturation Rydberg density depends
only very weakly on the density of ground state atoms $n_g$. In this
Letter we investigate the excitation dynamics across the condensation
temperature. Due to the bimodal density distribution in a
partially condensed cloud the time scales and saturation Rydberg
fractions are different in the condensate and thermal component. Figure
\ref{Fig_Rydberg_fraction}(a) shows the Rydberg fraction after
different excitation times as a function of temperature. The lines
are the result of a smoothing intended to guide the eye to the main
features of the measurement: At medium excitation times (\unit[320 and
370]{ns}) the Rydberg fraction decreases with proceeding formation of
the condensate i.e. reduced temperature. For the shortest and longest
excitation times, no such kink in the Rydberg fraction is visible at
the critical temperature.

The excitation times in this experiment are chosen such that for the
shortest time (170 ns), both components are just slightly blocked but
not saturated. For longer excitation times, the excitation is already
saturated in the condensate component while it is not yet in the
thermal part. For the longest excitation time (1970 ns), the excitation is
saturated in both components. As a start the following intuitive
picture may be used to explain the reduction of the total Rydberg
fraction at medium excitation times: With increasing BEC fraction,
more atoms are condensed into the BEC where the Rydberg fraction is
significantly lower than in the thermal cloud due to its higher
density, this also lowers the total Rydberg fraction. This effect
dominates over the smaller increase in the Rydberg fraction in the
thermal cloud as its density decreases (see
Fig.\,\ref{Fig_density_ratio}(a)). A deeper comprehension is obtained
by the following model based on collective states.

In this experiment the van der Waals interaction among the Rydberg
atoms blocks further excitation within a blockade radius $r_b$ around
one Rydberg atom \cite{Tong:2004,Singer:2004,Heidemann:2007}. This
limitation to one excitation per blockade sphere leads to a collective
Rabi oscillation which is enhanced in frequency by a factor
$\sqrt{N}$. In this sense the $N$ atoms per blockade sphere act like a
`superatom' \cite{Vuletic:2006} with a $\sqrt{N}$ larger transition
matrix-element. In the regime where the size of the sample is larger
than the blockade radius $r_{b}$, quantum correlations interconnect
the whole sample. Thus the system with $N_g\approx10^7$ atoms and
$N_R\approx10^3$ excitations is inaccessible to direct ab initio
simulations. The model described below reduces the quantum
correlations to spatial correlations and locally collective states by
assuming $N_{R}$ independent superatoms which oscillate at their
respective collective Rabi frequencies $\sqrt{N}\Omega_0$. We will
assume the superatoms to arrange in the close packing of a
face-centered cubic structure with 12 next neighbours. In this
hypothetical lattice, the saturation density is
$n_{R}=\sqrt{2}\,r_{b}^{-3}$ and the interaction energy at the center
of one superatom is $Z=14.5$ times the pairwise van der Waals ernergy.
$Z$ is slightly larger than 12 due to the atoms beyond the shell of
next neighbours. In a local density approximation the local number of
ground state atoms forming one superatom is given by
$N(r)=n_{g}(r)/n_{R}(r)$ and density variations on the scale of $r_b$
are neglected. The saturation density of Rydberg atoms can be derived
from the blockade condition which equates the van der Waals
interaction energy with the collective Rabi frequency, where $\kappa$
is a constant expected to be on the order of one:
\begin{equation}\label{eqn_blockade}\frac{Z\,C_6}{r_{b}^6(r)}=Z\,C_6\,\frac{1}{2}\,n_{R}^2(r)=\kappa\,\hbar\sqrt{N(r)}\,\Omega_0.
\end{equation}
The $C_6$-coefficient for the $43S$-state is $-1.7\times10^{19}$
a.u. \cite{Singer:2005}. As shown later, the blockade radius is
in the micrometer regime, substantially larger than the mean
interatomic distance. At these interatomic separations, the van der
Waals interaction that is used in Eq.~(\ref{eqn_blockade}) is
justified. From this relation, the local saturation density $n_R(r)$
and then the collective Rabi frequency $\Omega_c$ can be directly
derived.
\begin{eqnarray}\label{eqn_integration1}
n_{R}(r)&=&\left(\nicefrac{2\,\kappa\,\hbar}{Z\,C_6}
\right)^{\nicefrac{2}{5}}n_{g}^{\nicefrac{1}{5}}(r)\Omega_0^{\nicefrac{2}{5}},\\
\label{eqn_integration2}\Omega_{c}(r)&=&\sqrt{N(r)}\,\Omega_0=\left(\nicefrac{Z\,C_6}{2\,\kappa\,\hbar}
\right)^{\nicefrac{1}{5}}
n_{g}^{\nicefrac{2}{5}}(r)\Omega_0^{\nicefrac{4}{5}}
\end{eqnarray}

\begin{figure}
\includegraphics[width=86mm]{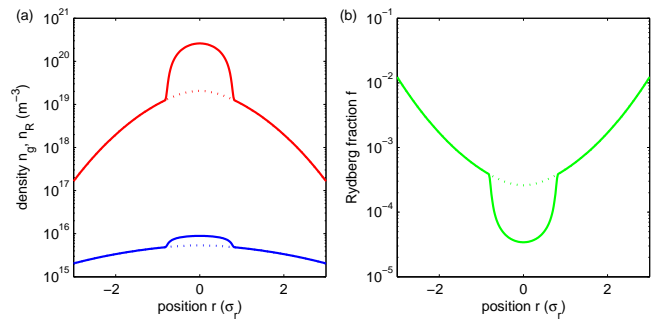}
\caption{\label{Fig_simulation_distribution} In (a) the radial cross sections
  of the density distributions are shown for the ground state atoms
  $n_g(r)$ in red/light grey and for the simulated saturation density
  of Rydberg atoms $n_R(r)$ in blue/dark grey. The cross section of
  the local Rydberg fraction $f(r)$ is shown in (b). Each figure shows
  the calculated distributions for the lowest temperature of
  Fig.\,~\ref{Fig_Rydberg_fraction} and the dotted lines represent
  the thermal component without BEC. As can be seen, $f(r)$ decreases with
  $n_g(r)$. Therefore in the BEC $f$ is substantially lower than in
  the thermal cloud.}
\end{figure}

To illustrate the calculated bimodality in the saturation density
of Rydberg atoms $n_R(r)$, Fig.\,\ref{Fig_simulation_distribution}
shows the density distributions as well as the distribution of the
fraction $f(r)$ which shows a significant reduction at the position of
the BEC component. The time dependence of the total number
$N_{R}(\tau)$ of Rydberg atoms can be obtained by integrating the
oscillations of the superatoms over the sample:
\begin{equation}
\label{eqn_integration3}N_{R}(\tau)=\int n_{R}(r) \cdot\sin^2 \left(\Omega_c(r)\,\tau/2 \right) d^3r
\end{equation}
The simulated behaviour of $N_R(\tau)$ nicely reproduces the
saturation curves shown in Ref.\,\cite{Heidemann:2007}. The derived
scaling of $N_{\text{sat}}\propto n_R\propto
n_g^{\nicefrac{1}{5}}\Omega_0^{\nicefrac{2}{5}}$ with Rabi frequency
is in excellent agreement with our previous observation where we found
$N_{\text{sat}}\propto
n_g^{0.07\pm0.02}\Omega_0^{0.38\pm0.04}$\cite{Heidemann:2007}, the
scaling with the peak density $n_g$ is close. The calculated scalings
of the initial slope of the excitation curve, $R\propto
n_R\,\Omega_{c}\propto
n_g^{\nicefrac{3}{5}}\Omega_0^{\nicefrac{6}{5}}$, are in good
agreement with the previous observation of $R\propto
n_g^{0.49\pm0.06}\Omega_0^{1.1\pm0.1}$ \cite{Heidemann:2007}.  This
simplified model, despite its approximate character, explained well the
observed scalings. Here we apply it for the situation of a changing
density distribution $n_g(r)$ upon condensation. The lines in
Fig.\,\ref{Fig_Rydberg_fraction}(b) show the calculated values for the
Rydberg fraction using this model. The atomic density distributions
were calculated from the information obtained from the absorption
images and averaged over several pictures taken at different
excitation times. The simulation uses two free parameters which is the
propotionality factor $\kappa$ in Eq.~(\ref{eqn_blockade}) and the
Rabi-frequency $\Omega_0$. The factor $\kappa$ was adjusted to 0.3
such that the measurement is best reproduced for high
temperatures. The Rabi frequency had to be reduced by a factor of 5.5
to reproduce the characteristic kink at the right excitation
times. This reduction is reasonable as the frequency of the excitation
lasers is only stable within 1.5 MHz between adjacent excitations thus
effectively reducing the Rabi-frequency in this model.  Note that the
simulation was used beyond the large radius $r$, where the model gets
unphysical due to $n_{R}(r)>n_{g}(r)$. For our parameters the effect
on the total fraction $f$ is negligible as only a fraction of at most
$10^{-4}$ of the atoms are at these large distances where the
collective behaviour ends and their single-atom excitation fraction is
negligible.

The scaling of the Rydberg fraction $f= N_R/N_g$ with density can be
derived from the model for short, moderate and long excitation times:
For short excitation times, when the dynamics cannot yet be
distinguished from single-atom behaviour, the Rydberg atom number is
proportional to $N_g$, therefore $f$ is independent of $n_g$. For
intermediate excitation times $\tau$, during the linear increase of
$N_{R}(\tau)$, $f$ can be approximated by $R\,\tau/N_g \propto
n_R\,\Omega_c\tau/n_g\propto n_g^{-\nicefrac{2}{5}}$. For long
excitation times, in the saturation regime, the dependence of the
fraction on the density is stronger: $f\propto
N_{\text{sat}}/N_g\propto n_R/n_g\propto
n_{g,0}^{-\nicefrac{4}{5}}$. These considerations hold for a change in
density without changing the shape of the distribution i.e. without
crossing the critical temperature or for the thermal component alone.

The simulation is able to reproduce the main features of the
experimental data: The dependence of $f$ on the density is increasing
for increasing excitation time.  For intermediate
excitation times, $f$ bends down below the condensation
temperature. At long excitation times, this kink is covered by the
increased dependence of $f$ on $n_g$ in the saturated thermal
cloud. To emphasize the effect of the bimodal density distribution,
the dashed lines in Fig.\,\ref{Fig_Rydberg_fraction}(b) show the
calculation which includes only the thermal component of the atomic cloud which
does not reproduce the characteristic kink.  According to this simulation,
the maximum density of Rydberg atoms in the BEC is \unit[9$\times
10^{15}$]{m$^{-3}$}, corresponding to a blockade radius of
\unit[5.4]{$\mu$m} and 5 Rydberg excitations within the volume of the
BEC.

The experimental fraction $f$ shows an overall increase with decreasing
temperature which is not sufficiently reproduced by the simulation. We
attribute this remaining discrepancy to a failure of the local
density approximation (LDA). The radial width of the gaussian
distribution gets as narrow as \unit[2]{$\mu$m} while the dynamics
depends on a density that is averaged over the size of the
blockade radius. For low temperatures this averaging effectively
lowers the ground state densities and leads to higher Rydberg
fractions than the simulation suggests.

Note that we do not expect the change in the density-density
correlation upon condensation to have a significant effect on the
described measurements as the de Broglie wavelength which is the
length scale on which the bunching in the thermal cloud occurs is on
the order of ($\lesssim$ \unit[350]{nm}) and therefore much smaller
than the blockade radius.


To conclude, we have demonstrated the Rydberg excitation of
Bose-Einstein condensates and show that the momentum distribution of
the condensate is not significantly affected by the presence of the
Rydberg atoms. We present measurements on the excitation dynamics
which show a significantly smaller Rydberg fraction in the condensed
sample than in the thermal sample. A simplified superatom model
reproduces the main features of the measurement as a consequence of
the changing density distribution upon condensation. The coherence of
the condensate was not relevant for the understanding of the
measurement of the overall excitation dynamics. Future studies will
focus on making use of the coherence properties of the condensate,
e.g. in order to measure the spatial density-density correlation
function of the Rydberg atoms by means of matter wave interferometry.

We acknowledge fruitful discussions with F. Robicheaux,
H.P. B\"{u}chler and L. Santos as well as financial support from the
Deutsche Forschungsgemeinschaft within the SFB/TRR21 and under the
contract PF 381/4-1, U.R. acknowledges support from the
Landesgraduiertenf\"{o}rderung Baden-W\"{u}rttemberg.\\


\begin{thebibliography}{10}
\providecommand{\bibnamefont}[1]{#1}
\providecommand{\bibfnamefont}[1]{#1}
\providecommand{\bibinfo}[2]{#2}

\bibitem{Stuhler:2005}
\bibinfo{author}{\bibfnamefont{J.}~\bibnamefont{Stuhler}}, \emph{et~al.},
  \bibinfo{journal}{Phys. Rev. Lett.} \textbf{\bibinfo{volume}{95}},
  \bibinfo{pages}{150406} (\bibinfo{year}{2005}).

\bibitem{Low:2005}
\bibinfo{author}{\bibfnamefont{R.}~\bibnamefont{L{\"o}w}},
  \bibinfo{author}{\bibfnamefont{R.}~\bibnamefont{Gati}},
  \bibinfo{author}{\bibfnamefont{J.}~\bibnamefont{Stuhler}}, \bibnamefont{and}
  \bibinfo{author}{\bibfnamefont{T.}~\bibnamefont{Pfau}},
  \bibinfo{journal}{Europhys. Lett} \textbf{\bibinfo{volume}{71}},
  \bibinfo{pages}{214} (\bibinfo{year}{2005}).

\bibitem{O'Dell:2000}
\bibinfo{author}{\bibfnamefont{D.}~\bibnamefont{O'Dell}},
  \bibinfo{author}{\bibfnamefont{S.}~\bibnamefont{Giovanazzi}},
  \bibinfo{author}{\bibfnamefont{G.}~\bibnamefont{Kurizki}}, \bibnamefont{and}
  \bibinfo{author}{\bibfnamefont{V.~M.} \bibnamefont{Akulin}},
  \bibinfo{journal}{Phys. Rev. Lett.}
  \textbf{\bibinfo{volume}{84}}(\bibinfo{number}{25}), \bibinfo{pages}{5687}
  (\bibinfo{year}{2000}).

\bibitem{Santos:2000}
\bibinfo{author}{\bibfnamefont{L.}~\bibnamefont{{Santos}}},
  \bibinfo{author}{\bibfnamefont{G.~V.} \bibnamefont{{Shlyapnikov}}},
  \bibinfo{author}{\bibfnamefont{P.}~\bibnamefont{{Zoller}}}, \bibnamefont{and}
  \bibinfo{author}{\bibfnamefont{M.}~\bibnamefont{{Lewenstein}}},
  \bibinfo{journal}{Phys. Rev. Lett.} \textbf{\bibinfo{volume}{85}},
  \bibinfo{pages}{1791} (\bibinfo{year}{2000}).

\bibitem{Greene:2000}
\bibinfo{author}{\bibfnamefont{C.~H.} \bibnamefont{Greene}},
  \bibinfo{author}{\bibfnamefont{A.~S.} \bibnamefont{Dickinson}},
  \bibnamefont{and} \bibinfo{author}{\bibfnamefont{H.~R.}
  \bibnamefont{Sadeghpour}}, \bibinfo{journal}{Phys. Rev. Lett.}
  \textbf{\bibinfo{volume}{85}}(\bibinfo{number}{12}), \bibinfo{pages}{2458}
  (\bibinfo{year}{2000}).

\bibitem{Greene:2006}
\bibinfo{author}{\bibfnamefont{C.~H.} \bibnamefont{Greene}}, \emph{et~al.},
  \bibinfo{journal}{Phys. Rev. Lett.}
  \textbf{\bibinfo{volume}{97}}, \bibinfo{eid}{233002}
  (\bibinfo{year}{2006}).

\bibitem{Chikkatur:2000}
\bibinfo{author}{\bibfnamefont{A.~P.} \bibnamefont{{Chikkatur}}},
  \emph{et~al.}, \bibinfo{journal}{Phys. Rev. Lett.}
  \textbf{\bibinfo{volume}{85}}, \bibinfo{pages}{483} (\bibinfo{year}{2000}).

\bibitem{Massignan:2005}
\bibinfo{author}{\bibfnamefont{P.}~\bibnamefont{{Massignan}}},
  \bibinfo{author}{\bibfnamefont{C.~J.} \bibnamefont{{Pethick}}},
  \bibnamefont{and} \bibinfo{author}{\bibfnamefont{H.}~\bibnamefont{{Smith}}},
  \bibinfo{journal}{\pra} \textbf{\bibinfo{volume}{71}}(\bibinfo{number}{2}),
  \bibinfo{pages}{023606} (\bibinfo{year}{2005}).

\bibitem{Vogt:2007}
\bibinfo{author}{\bibfnamefont{T.}~\bibnamefont{{Vogt}}}, \emph{et~al.},
  \bibinfo{journal}{Phys. Rev. Lett.}
  \textbf{\bibinfo{volume}{99}}(\bibinfo{number}{7}), \bibinfo{pages}{073002}
  (\bibinfo{year}{2007}).

\bibitem{Buchler:2007}
\bibinfo{author}{\bibfnamefont{H.~P.} \bibnamefont{{B{\"u}chler}}},
  \bibinfo{author}{\bibfnamefont{A.}~\bibnamefont{{Micheli}}},
  \bibnamefont{and} \bibinfo{author}{\bibfnamefont{P.}~\bibnamefont{{Zoller}}},
  \bibinfo{journal}{Nature Physics} \textbf{\bibinfo{volume}{3}},
  \bibinfo{pages}{726} (\bibinfo{year}{2007}).

\bibitem{Loew:2007}
\bibinfo{author}{\bibfnamefont{R.}~\bibnamefont{Loew}}, \emph{et~al.},
  \bibinfo{journal}{eprint arXiv.org:0706.2639}  (\bibinfo{year}{2007}).

\bibitem{CommentRabifrequency:2007}
\bibinfo{note}{This value is estimated from our calculation of the dipole
  matrix element, which is in excellent agreement with an independent
  calculation of F. Robicheaux (priv. comm.)}.

\bibitem{Heidemann:2007}
\bibinfo{author}{\bibfnamefont{R.}~\bibnamefont{Heidemann}}, \emph{et~al.},
  \bibinfo{journal}{Phys. Rev. Lett.}
  \textbf{\bibinfo{volume}{99}}, \bibinfo{eid}{163601}
  (\bibinfo{year}{2007}).

\bibitem{Clark:1954}
\bibinfo{author}{\bibfnamefont{P.~J.} \bibnamefont{Clark}} \bibnamefont{and}
  \bibinfo{author}{\bibfnamefont{F.~C.} \bibnamefont{Evans}},
  \bibinfo{journal}{Ecology} \textbf{\bibinfo{volume}{35}},
  \bibinfo{pages}{445} (\bibinfo{year}{1954}).

\bibitem{commentlowess:2007}
\bibinfo{note}{The smoothing was done with a robust locally weighted linear
  regression method (lowess), using the 50\% nearest data points at a time for
  regression \cite{Cleveland:1979}.}

\bibitem{Tong:2004}
\bibinfo{author}{\bibfnamefont{D.}~\bibnamefont{{Tong}}}, \emph{et~al.},
  \bibinfo{journal}{Phys. Rev. Lett.}
  \textbf{\bibinfo{volume}{93}}(\bibinfo{number}{6}), \bibinfo{pages}{063001}
  (\bibinfo{year}{2004}).

\bibitem{Singer:2004}
\bibinfo{author}{\bibfnamefont{K.}~\bibnamefont{{Singer}}}, \emph{et~al.},
  \bibinfo{journal}{Phys. Rev. Lett.}
  \textbf{\bibinfo{volume}{93}}(\bibinfo{number}{16}), \bibinfo{pages}{163001}
  (\bibinfo{year}{2004}).

\bibitem{Vuletic:2006}
\bibinfo{author}{\bibfnamefont{V.}~\bibnamefont{Vuletic}},
  \bibinfo{journal}{Nature Physics} \textbf{\bibinfo{volume}{2}},
  \bibinfo{pages}{801} (\bibinfo{year}{2006}).

\bibitem{Singer:2005}
\bibinfo{author}{\bibfnamefont{K.}~\bibnamefont{{Singer}}},
  \bibinfo{author}{\bibfnamefont{J.}~\bibnamefont{{Stanojevic}}},
  \bibinfo{author}{\bibfnamefont{M.}~\bibnamefont{{Weidem{\"u}ller}}},
  \bibnamefont{and}
  \bibinfo{author}{\bibfnamefont{R.}~\bibnamefont{{C{\^o}t{\'e}}}},
  \bibinfo{journal}{J. Phys. B: At., Mol. Opt. Phys.}
  \textbf{\bibinfo{volume}{38}}, \bibinfo{pages}{S295} (\bibinfo{year}{2005}).

\bibitem{Cleveland:1979}
\bibinfo{author}{\bibfnamefont{W.}~\bibnamefont{Cleveland}},
  \bibinfo{journal}{J. Amer. Statist. Assoc.}
  \textbf{\bibinfo{volume}{74}}(\bibinfo{number}{368}), \bibinfo{pages}{829}
  (\bibinfo{year}{1979}).

\end{thebibliography}

\end{document}